\documentclass[aps,prl,twocolumn,floatfix,footinbib,showpacs,superscriptaddress]{revtex4-1}
\usepackage[english]{babel}
\usepackage[utf8]{inputenc}
\usepackage{graphicx}
\usepackage{fancyhdr}
\usepackage[active]{srcltx}
\usepackage{setspace}
\usepackage{float}
\usepackage{amsmath,bbm}
\usepackage{amsfonts}
\usepackage{natbib}
\usepackage{color}
\usepackage{hyperref}

\begin{document}

\title{Sea of Majorana fermions from pseudo-scalar superconducting order in three dimensional Dirac materials}

\author{Morteza Salehi}
\affiliation{Department of Physics, Sharif University of Technology,Tehran 11155-9161, Iran}
\author{S. A. Jafari}
\affiliation{Department of Physics, Sharif University of Technology,Tehran 11155-9161, Iran}
%\affiliation{Center of Excellence for Complex Systems and Condensed Matter (CSCM), Sharif University of Technology, Tehran 1458889694, Iran}
\affiliation{Theoretische Physik, Universit\"at Duisburg-Essen, 47048 Duisburg, Germany}

%===============================================================================

\begin{abstract}
We find that {\em singlet} superconducting pairing can lead to Majorana fermions in three dimensional Dirac superconductors (3DDS) if the 
pairing order parameter is a pseudo-scalar, i.e. it changes sign under mirror reflection. The pseudo-scalar
superconducting order parameter, $\Delta_5$ can close and reopen the spectral gap caused by the scalar Dirac 
mass $m$ in a three-dimensional Dirac material (3DDM), giving rise to a two-dimensional Majorana sea (2DMS) at
the plane of the gap kink. By bringing the Hamiltonian into a canonical form which then gives the winding number, 
we show that this system belongs to the 
DIII class of topological superconductors. We calculate the transport signature of 2DMS, namely a perfect
Andreev-Klein transmission that manifests in a robust peak in the differential conductance. Further, we find the
$4\pi$ periodicity in the $\Delta_5|m|\Delta_5$ Josephson junctions. Gauge transformed version of the present scenario
implies that the interface of a conventional s-wave superconductor with a peculiar 3DDM whose mass is a pseudo-scalar,
$m_5$ also hosts a 2DMS. 
\end{abstract}

\pacs{}

\maketitle

%===============================================================================

\textit{Introduction}.-- Band topology in insulators and superconductors is connected with the  change in the sign of the gap parameter which in turn gives rise to zero energy states at the location of gap kink. This mechanism in the case of insulators gives rise to gapless surface modes protected by a topological invariant~\cite{BernevigBook,Zhang2011RMP,ShenBook2013}. When the spectral gap is of the superconducting (pairing) 
nature, these topologically protected modes will be Majorana zero modes, which are their own 
anti-particles~\cite{Alicea,Beenakker2013rev}. To realize Majorana fermions (MFs) various scenarios have been proposed 
which involve closing and re-opening the superconducting gap in one way or another. Gaping chiral modes of topological insulators (TIs) by Zeeman field and superconducting pairing gives rise to MFs in the interface region where these two 
gapping mechanisms are of comparable strength~\cite{Fu2008PRL, Fu2009PRB}. In one-dimensional nano-wires this can be achieved by the competition between a polarizing Zeeman field, and depolarizing spin-orbit 
interaction~\cite{Sau,vonOppen}. Alternative scenario is to generate MFs involve driving supercurrent through a superconductor (SC) next to a TI in a pairing dominated gap regime where twist due to supercurrent reduces the phase space for Cooper pairing and gives way to Zeeman dominated gap~\cite{Romito}. 
In two-dimensions, the vortex core of a p-wave SC binds a MF~\cite{Stone2004}. The required p-wave SC can be engineered on the surface of a TI~\cite{Fu2008PRL} by proximity to a conventional s-wave SC. 
The above scenarios are first of all limited to low dimensions, and secondly require a TR breaking by a Zeeman field. 

In this letter we propose yet another {\em three dimensional} system that admits a two dimensional sheet of MFs 
without requiring a Zeeman field. We find that a peculiar {\em pseudo-scalar} (odd-parity) superconducting order 
parameter, $\Delta_5$ can give rise to a two-dimensional sheet of MFs when it is interfaced with a 
three-dimensional Dirac material (3DDM).
To set the stages for our finding, let us start by noting that
in a one-band situation described by a parabolic band dispersion, the strength of the gap is characterized by a (scalar) 
gap parameter. However in 3DDM where the relevant degrees of freedom are described by the Dirac equation~\cite{Fuseya2011PRL}, new possibilities can arise.
We assume a 3DDM which has a single Dirac point in the $\Gamma$ point of it's Brillouin zone ~\cite{Zhang2009NP,Narayan2014PRL,Liu2014Science}.
For such a 3DDM, the low-energy degrees of freedom are described by 4-component Dirac spinors 
and hence the most general pairing is 
$\bar\psi\hat{\Delta}_S\psi_c=\psi^\dagger\gamma^0\hat{\Delta}_S\psi_c$
where $\psi_c$ satisfies the same Dirac equation as $\psi$, but with opposite charge and
is covariant under Lorentz transformation. 
%The explicit form  of $\psi_c$ in terms of $\psi$ depends on the basis used to construct the Clifford algebra
%and will be specified in the text. 
The $4\times 4$ pairing matrix $\hat{\Delta}_S$ can be expanded in terms of a basis composed of $\mathbbm{1}$, 
four $\gamma^\mu$, $\gamma^5$, four $\gamma^5\gamma^\mu$ and six anti-commutators 
$\sigma^{\mu\nu}=i[\gamma^\mu,\gamma^\nu]/2$~\cite{ZeeBookQFT} as 
$\hat{\Delta}_S=\Delta_0\mathbbm{1}+\Delta_\mu\gamma^\mu+\Delta_5\gamma^5+\Delta_{5\mu}\gamma^5\gamma^\mu+\Delta_{\mu\nu}\sigma^{\mu\nu}$. 
Then the pairing $\Delta_0$ will be the conventional scalar pairing, while $\Delta_5$ will be a pseudo-scalar 
pairing under the Lorentz transformation. Similarly $\Delta_\mu,\Delta_{5\mu}, \Delta_{\mu\nu}$, will be vector, 
pseudo-vector and tensor superconducting pairings. We have examined all the above $16$ pairing channels and find that
the pseudo-scalar superconducting gap $\Delta_5$ works in an opposite direction to the normal Dirac gap $m\gamma^0$.
Therefore placing a $m$ dominated region (i.e. a 3DDM) next to a $\Delta_5$ dominated region denoted as $m|\Delta_5$,
gives rise to a two-dimensional sheet of Majorana fermions. This mechanism, {\em does not require p-wave pairing, nor 
a magnet to break time-reversal}. Instead, it requires a peculiar form of pseudo-scalar
superconducting order parameter that breaks the mirror symmetry. 
We then proceed to show that such a system belongs to the  DIII topological class allowing
$Z$-number classification which in turn guarantees the existence of two-dimensional Majorana sea (2DMS) 
on the region where the strength of $m$ and $\Delta_5$ are equal. We further corroborate our results by
showing a perfect Andreev reflection and robust zero-mode resonance peak in differential conductance and
fractional supercurrent in $\Delta_5|m|\Delta_5$ Josephson junction. 

\begin{figure}[t]
\includegraphics[scale=0.3]{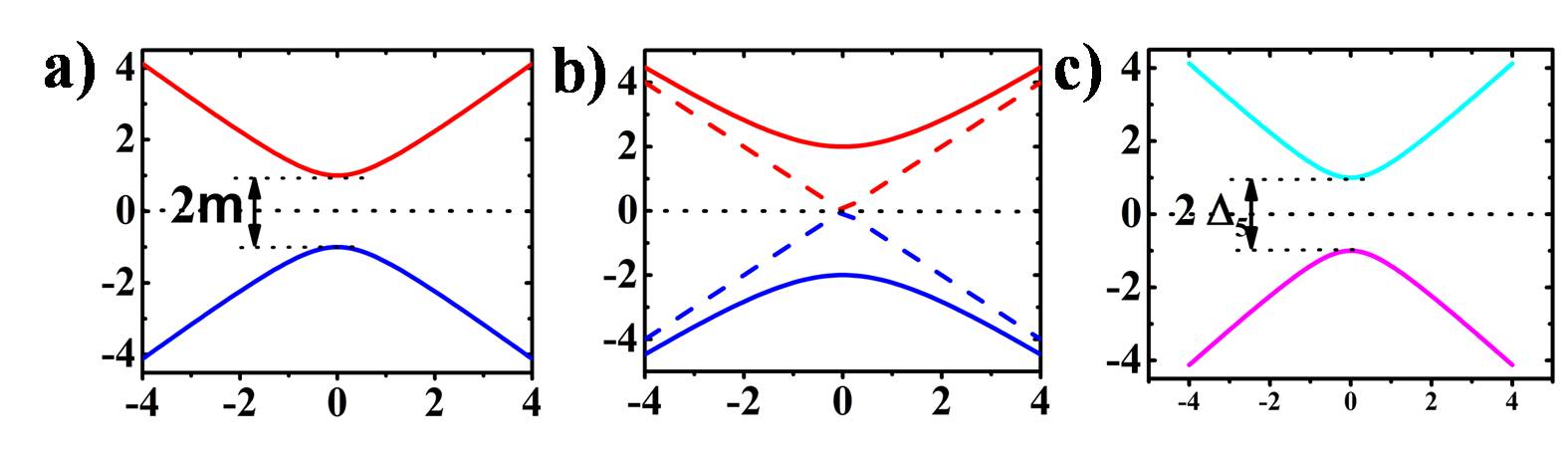}
\caption{(Color online) (a) Dispersion relation of 3DDM regime. (b) Gap closing and topological phase transition 
for one band whereas the other one remains trivial. (c) Dispersion relation of 3DDS regime. In these results we 
set $\mu=0$. The length scale $l_D=\hbar/ m v_F$ is set by the energy gap of the 3DDM.}
\label{Fig.TopologicalPhaseTransition}
\end{figure}

{\em Model:} We consider a Dirac material with a single Dirac cone,
\begin{equation}
\mathcal{H}(\textbf{k})= v_F \gamma_0\left( \hbar \boldsymbol{\gamma}.\mathbf{k}+ m v_F\right).
\label{Eq.01}
\end{equation} 
We use the representation $\gamma^0=\tau_3 \otimes \sigma_0$ and $\gamma^j=\tau_1 \otimes i \sigma_j$
for the Clifford algebra where $\sigma_j$ and $\tau_j$ are Pauli matrices acting on spin and band spaces, 
respectively. Also, $\sigma_0$ and $\tau_0$ are the $2 \times2$ unit matrices.
The $\textbf{k}$ is the wave vector of excitations. In four space-time dimensions one can also 
construct $\gamma^5=i\gamma^0\gamma^1\gamma^2\gamma^3$ which will be very essential for our discussion in this
paper. In Eq.(\ref{Eq.01}), the mass term is of the ordinary $m\gamma^0$ nature 
and is responsible for the band gap, and $v_F$ is the Fermi velocity. 
The pairing Hamiltonian for such a system is:
\begin{equation}
H_{\textit{BCS}}=\frac{1}{2}\int d\textbf{r}\left(\begin{array}{cc}
\psi^\dagger & \psi_c^\dagger
\end{array}\right)H_{\rm DBdG}\left(\begin{array}{c}
\psi \\
\psi_c
\end{array}\right),
\label{Eq.BCS}
\end{equation}
corresponding to which the Dirac-Bogoliubov-deGennes (DBdG) equation in \textbf{k}-space is,
\begin{equation}
\left( \begin{array}{cc}
\mathcal{H}(\textbf{k})-\mu & \gamma_0\hat{\Delta}_S e^{i \phi}\\
\hat{\Delta}^\dagger_S\gamma_0 e^{-i \phi} & \mu+\mathcal{C}\mathcal{H}(\textbf{k})\mathcal{C}^{-1} \\
\end{array}\right)
\left( \begin{array}{c}
u \\
v\\
\end{array}\right)= \varepsilon \left(\begin{array}{c}
u \\
v \\
\end{array} \right),
\label{Eq.DBdG}  
\end{equation}
where $\varepsilon$ is the energy of eigenstate with respect to the chemical potential $\mu$. 
Here $\phi$ is the macroscopic phase of superconductor, $u  (v)$  is the 
electron (hole) part of BdG wave function in Nambu space. The anti-unitary operator $\mathcal{C}=\gamma_2\gamma_0 K$ is 
the charge-conjugation of Eq.(\ref{Eq.01}), where $K$ is the complex-conjugation: 
%This means that 
%if $\psi$ satisfies the Dirac equation for an electrons, then $\psi_c={\cal C}\gamma^0\psi=\gamma^2\psi^*$ satisfies 
%the Dirac equation with an opposite charge and transforms under the Lorentz transformation the same way as $\psi$ 
%does~\cite{ZeeBookQFT}. 
This means that Lorentz transformation for $\bar\psi_c=\psi_c^\dagger\gamma^0$ 
is the inverse of the Lorentz transformation for $\psi$. Therefore requiring the superconducting pairing to be
Lorentz invariant~\cite{ZeeBookQFT,Capelle1999PRB2}, the scalar (conventional) superconducting pairing is given by 
$\bar\psi_c\Delta_s\mathbbm{1}\psi=\psi_c^\dagger\gamma^0\Delta_s\mathbbm{1}\psi\sim v\gamma^0\Delta_s u$. 
Similarly pseudo-scalar pairing 
is $\bar\psi_c\Delta_5\gamma^5\psi=\psi_c^\dagger \Delta_5\gamma^0\gamma^5\psi\sim v \Delta_5 \gamma^0\gamma^5 u$, etc.
We have examined all of the above 16 possible superconducting pairing channels. We find that only the $\Delta_5$ parameter
gives rise to a gap closing when both $m$ and $\Delta_5$ are present. 
To see this in $\Delta_5$ channel, let us for clarity of notation set $\hbar$ and $v_F$ to $1$ which 
gives the eigenvalues of Eq.~(\ref{Eq.DBdG}) as $\varepsilon=\pm\sqrt{A\pm2 \sqrt{B}}$, 
where $A=k^2+m^2+\mu^2+\Delta_5^2$ and $B=\Delta_5^2 m^2+(m^2+k^2)\mu^2$.
When the Dirac mass $m$ dominates the spectral gap ($\Delta_5\to 0$) the above eigenvalues 
reduce to $ \varepsilon=\pm \mu \pm\sqrt{k^2+\Delta_5^2}$.
In this case each eigenvalue is doubly (spin) degenerate~\cite{Supplement}. 
Deep in the 3DDS region where $\Delta_5$ dominates the eigenvalues 
acquire the following structure $\varepsilon=\pm\sqrt{(k\pm \mu)^2+\Delta_5^2}$
which corresponds to a pairing gap at the Fermi level. In a region where both $m$ and $\Delta_5$ are
non-zero the nature of gap can be more clearly seen if we look at $\mu=0$ where dispersion becomes
$\varepsilon=\pm\sqrt{k^2+(m\pm \Delta_5)^2}$ and the band gap is determined by
$\Delta_\pm=(m\pm\Delta_5)$ which clearly indicates the competition between the Dirac mass $m$ and
the pseudo-scalar superconducting parameter $\Delta_5$. 
Fig.~\ref{Fig.TopologicalPhaseTransition} summarizes the closing and reopening of the spectral gap
as one moves from $m$ dominated region to $\Delta_5$ dominated region.
The closing of the Dirac gap and re-opening of it in the form of a pseudo-scalar superconducting gap
in this system does not require any magnetic field. This implies that when a $\Delta_5$ 3DDS
is brought next to a normal 3DDM with gap parameter $m$, such that $\Delta_5>m$, the induced $\Delta_5$ on
the normal 3DDM side decays towards the bulk of 3DDM and crosses $m$ somewhere in the interface
where the excitations become gapless. 

Let us now prove that when $\Delta_5=m$ a two dimensional Majorana sea appears.
The pseudo-scalar character of $\Delta_5$ pairing can be interpreted as a {\em singlet} 
superconductor whose mirror image has an opposite sign.
To understand further the properties of such a pairing, 
let us now explicitly construct the anti-unitary operators corresponding to particle-hole,
and time-reversal transformation for Eq.~\eqref{Eq.DBdG}:
Let $\eta_j$ be set of Pauli matrices in the Nambu space. Then the particle-hole 
and time-reversal symmetries in this space
can be defined as $PH=i\eta_2\otimes\gamma^0\gamma^5\gamma^2 K$ and
$TR=\eta_0\otimes \gamma^0\gamma^1\gamma^3K$, respectively.
Owing to $PH^2=1$ and $TR^2=-1$, the chiral symmetry $SL=PH*TR$ satisfies $SL^2=1$ which places the present system in the
DIII class of topological superconductors which can be classified with winding number.
We explicitely obtain the topoligical charge~\cite{Supplement},
\begin{equation}
%\mathcal{Q}=\frac{1}{2}\left(\frac{m+\Delta_5}{|m+\Delta_5|}
%-\frac{m-\Delta_5}{|m-\Delta_5|}\right).
{\cal Q}=\Theta(|\Delta_5|-|m|)~\text{sign}(\Delta_5)
\label{Eq.WindingNumber}
\end{equation}
which clearly shows that ${\cal Q}$ can be $0,\pm 1$, and hence
a $Z$ number classification~\cite{Schnyder2008PRB,Chiu2016RMP}.

\begin{figure}[b!]
\includegraphics[scale=0.25]{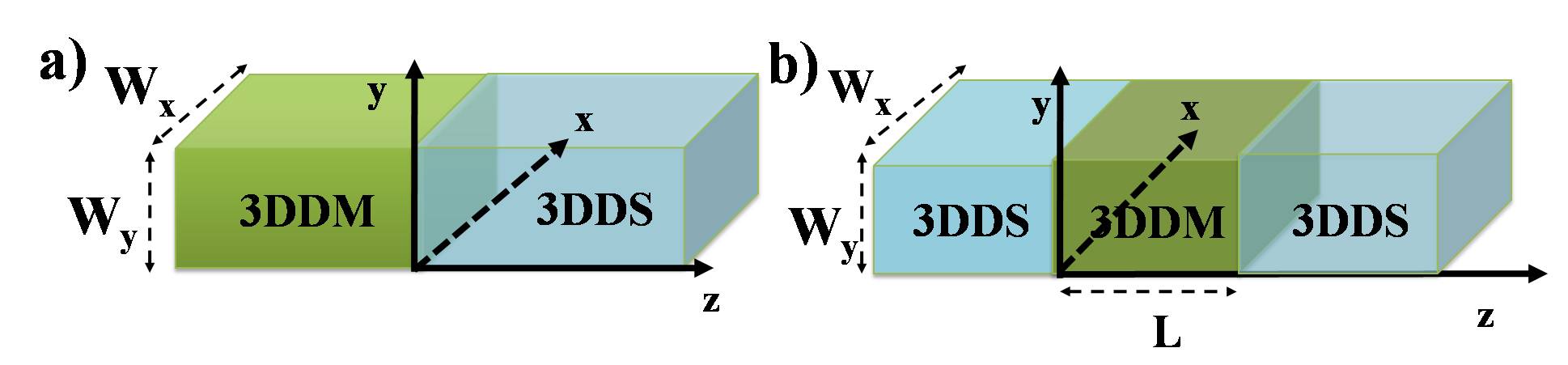}
\caption{(Color online) (a) Schematic illustration of 3DDM$|$3DDS junction ($m|\Delta_5$). 
We assume a 3DDM can be superconductor by proximity effect. (b) The Josephson junction of 
3DDS$|$3DDM$|$3DDS ($\Delta_5|m|\Delta_5$).}
\label{Fig.structure}
\end{figure}

\begin{figure}
\includegraphics*[width=6 cm, height=2.5 cm]{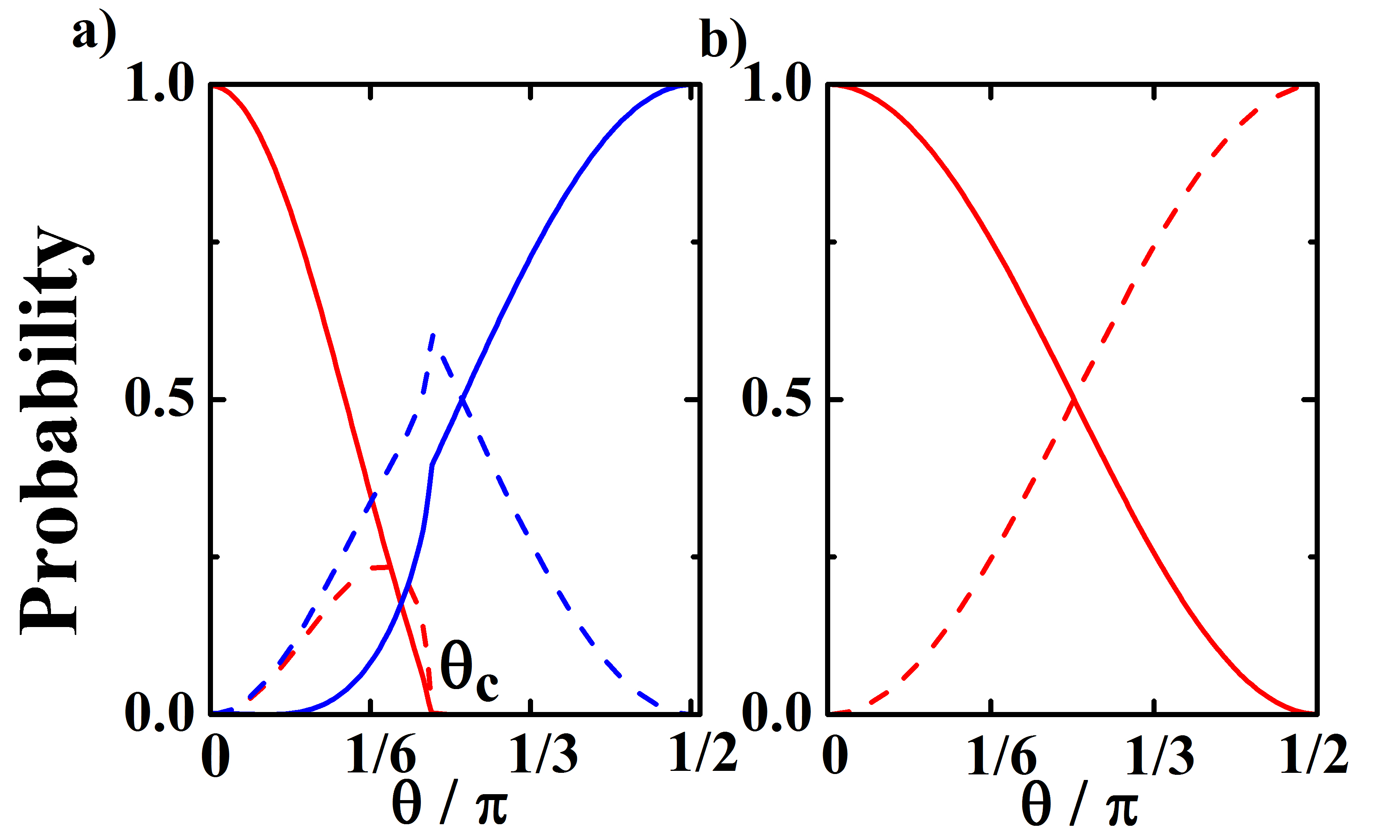}
\caption{(Color online) Probability of Andreev and normal reflections vs polar angle of incident. The conventional (spin-flipped) normal reflection is shown by solid (dashed) blue line. Also, the conventional (anomalous) Andreev reflection is prevaricated by solid (dashed)red line. (a) For $\varepsilon= \Delta_5$, the Andreev-Klein tunneling occurs in $\theta=0$. (b) In $\varepsilon=0$, the perfect Andreev reflection occurs for all angles of incidence. The input values are $ \mu=m, \Delta_5=5m$.}
\label{PvsAngle}
\end{figure}

\begin{figure}[bh]
\includegraphics[width=5cm, height=3.25cm]{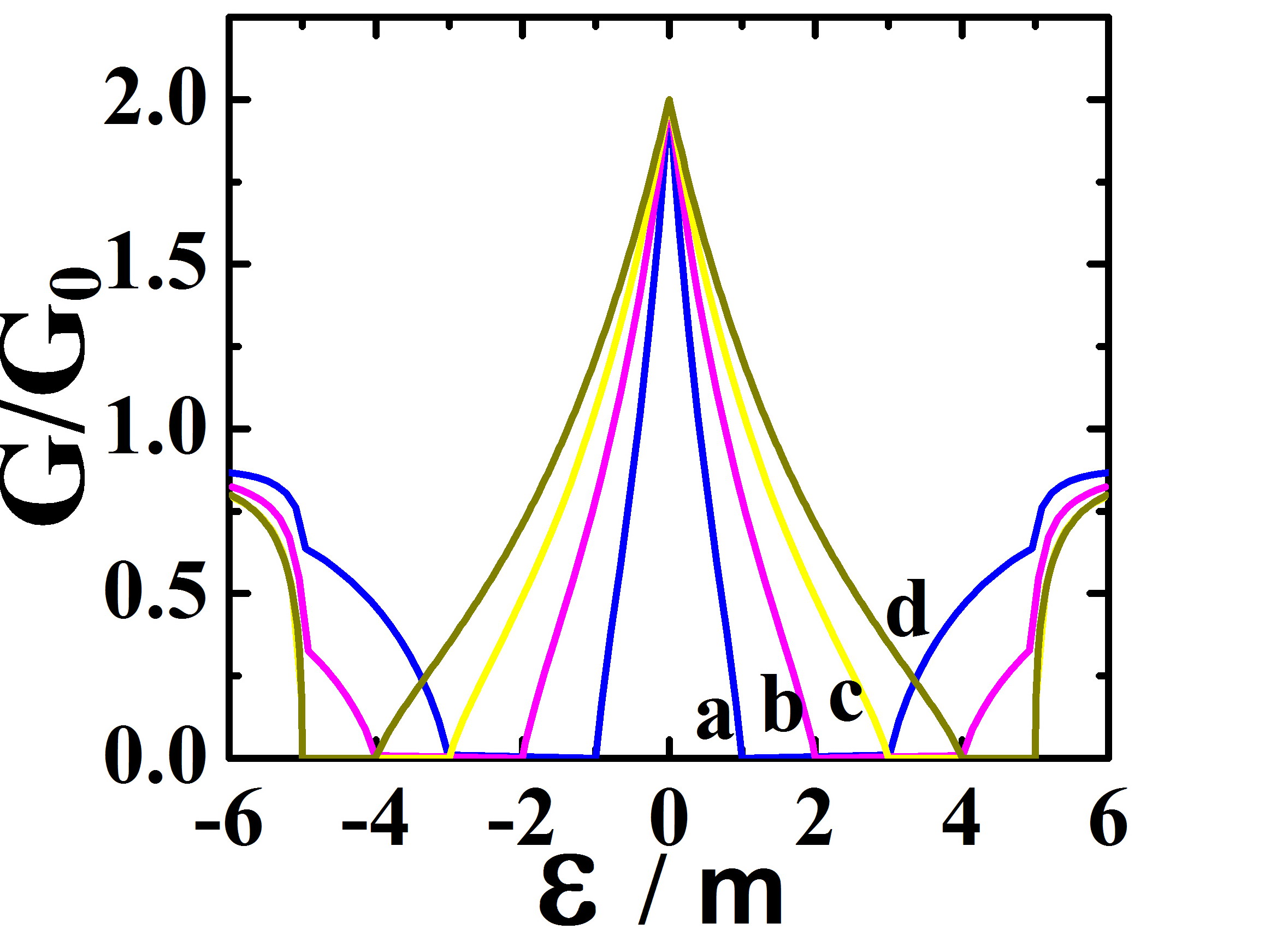}
\caption{(Color online) Conductance of the $m|\Delta_5$ system. 
We set $\Delta_5=5 m$ and $\mu/m=\{2,3,4,5\} $ for $\{a, b, c, d\}$ curves respectively.}
\label{Fig.conductance}
\end{figure}

{\em Transport signatures of Majorana sea}. To further corroborate our central result concerning a sea of Majorana fermions in the
present system, let us now focus on the transport properties arising from 2DMS which can be directly accessed in experiments. 
Let us begin by looking into a single $m|\Delta_5$ junction. The lateral coordinates in the 
plane of junction are $(x,y)$ confined within a lateral dimensions $(W_x,W_y)$, which means the corresponding components of wave vector in these directions are quantized as $k_{x(y)}=(n_{x(y)}+1/2)\pi/W_{x(y)}$. 
Each mode can be identified with a set of quantum number $n=(n_x,n_y)$.  In the 3DDM side of the junction ($z<0$), there are eight components of wave functions, $\Psi^{M,\pm}_{e(h) \uparrow(\downarrow)}$ where the indices $e$ ($h$) 
characterize electron (hole)-like quasi-particles, $\uparrow (\downarrow)$ denotes its spin direction with respect to $z$-axis 
and $\pm$ indicates the right or left mover character of carriers~\cite{Supplement}. 
The interface between $m$, ($z<0$) and $\Delta_5$ ($z > 0$) regions can reflect an incident electron as a hole. In this 
Andreev reflection process, the missing charge of $2e$ is absorbed by superconductor as Cooper pair~\cite{Andreev1964JETP}. 
Typically the s-wave assumption of superconductivity imposes that the reflected hole from an $\uparrow$-spin incident electron must be 
in a $\downarrow$-spin state and vice versa. However, 
due to strong spin-orbit coupling encoded in the Dirac Hamiltonian of a 3DDM, when an electron in a given 
spin direction hits a 3DDM, the spin of the electron transmitted into 3DDM can be flipped~\cite{Salehi2015AOP}. 
Therefore in 3DDM$|$3DDS junction we also take into account the possibility of
an anomalous -- i.e. spin flipped -- Andreev reflection of the reflected hole. The boundary condition for
an $\uparrow$ spin electron hitting the interface is given by,
\begin{equation}
\begin{array}{rl}
\Psi_{e,\uparrow}^{M,+} & +\sum_{\nu=\uparrow,\downarrow}r_{N,\alpha}\Psi_{e,\nu}^{M,-}+\sum_{\nu=\uparrow,\downarrow}r_{A,\nu}\Psi_{h,\nu}^{M,-}\\
& =\sum_{\kappa=1,2}t_{e,\kappa}\Psi_{e,\kappa}^{S,+}+\sum_{\kappa=1,2}t_{h,\kappa}\Psi_{h,\kappa}^{S,+}
\end{array}.
\label{BoundaryCondition}
\end{equation}
The left hand side is related to the wave function in 3DDM side, and the right hand side describes the wave equation in 3DDS side.
Here $r_{N,\uparrow}$ and $r_{N,\downarrow}$ are the amplitudes of conventional and spin-flipped normal reflection, respectively. 
The $r_{A,\downarrow}$ and $r_{A,\uparrow}$ are the amplitudes of conventional and anomalous Andreev reflection, respectively. 
Similar processes take place for $\downarrow$ spin and hole carriers as well~\cite{Supplement}. The probability of these reflections 
vs. $\theta$, the polar angle of incidence, are depicted in Fig.~(\ref{PvsAngle}). Because of the conservation of 
parallel component of wave vector, $k_n=\sqrt{k_x^2+k_y^2}$ at the scattering process, the angle of propagation for reflected 
hole ($\theta'$) has a critical value $\theta_C=\sin^{-1}((\mu-\varepsilon)^2-\Delta_D^2)/(\mu+\varepsilon)^2-\Delta_D^2))$  beyond which
the reflected hole can not contribute to transport. For zero modes, $\varepsilon= 0$, the conventional and spin-flipped 
normal reflections would disappear and we are left with conventional and anomalous Andreev reflection given by,
\begin{equation}
r_{A,\downarrow}=-e^{-i \phi} \cos\theta,~~~~
r_{A,\uparrow}= e^{i \alpha-i\phi}\sin\theta
\label{Eq.08zero-energyAR}
\end{equation}
 where, $\alpha=\arctan(k_y/k_x)$ is azimuthal angle. From Eq.~(\ref{Eq.08zero-energyAR}), it is obvious that for a zero-energy 
incident electron at any angle of propagation we have a perfect Andreev reflection, 
$|r_{A,\downarrow}|^2+|r_{A,\uparrow}|^2=1$.
This is a transport signature of Majorana fermions~\cite{Beenakker2013rev,Knez2012PRL}. This effect is robust against 
changing the chemical potential and angles of incidence ($\alpha, \theta$). 
The BTK formula for the differential conductance of the junction~\cite{Blonder1982PRB} will be,
\begin{equation}
   \frac{G(\varepsilon)}{G_0}=\sum_n\left[1+\sum_{\nu}\left|r_{A,\nu}(\varepsilon,k_n)\right|^2-\left|r_{N,\nu}(\varepsilon,k_n)\right|^2\right]
\label{Eq.Conductance}
\end{equation}
where $G_0=e^2/h $ is the quantum of conductance. Note that the summation over $\nu$ includes spin-flipped
contributions as well. 
When the linear dimensions of the interface are much larger than the superconducting coherence 
length, $\xi_s=\hbar v_F/\Delta_5$, i.e. $W_x,W_y\gg \xi_s$, the summation over mode indices $n$ 
in Eq.~(\ref{Eq.Conductance}) can be replaced by 
an integral. The conductance for various values of chemical potential has been plotted in Fig.~(\ref{Fig.conductance})
as a function of energy. Note that the height of the resonance peak at zero energy is pinned at $2G_0$ which indeed arises
from the 2DMS. 

Now let us move to the next transport signature of 2DMS, namely the fractional Josephson current. Consider the geometry
depicted in part(b) of Fig.~(\ref{Fig.structure}) and assume that $L$ is the length of junction and $\delta\phi=\phi_R-\phi_L$ is 
the phase difference between the two $\Delta_5$ superconductors. 
In th short junction limit, $L \ll \xi_S$, the Andreev bound states (ABS) 
responsible for carrying the super-current between two 3DDS region are given by~\cite{Supplement},
\begin{equation}
\epsilon_{\pm,n}(\delta\phi)=\pm\Delta_S\sqrt{\tau_n}\cos(\delta\phi/2).
\label{Eq.12}
\end{equation}
where $\tau_n$ is the normal transmission probability of the junction,
$\tau_n=\left(\cosh^2\kappa_nL+\mu^2\sinh^2\kappa_nL/\kappa_n^2\right)^{-1}$,
with $\kappa_n=\sqrt{\Delta_D^2+k_n^2-\mu^2}$. 
Despite that the Hamiltonian in Eq.~(\ref{Eq.DBdG}) is invariant under $2\pi$ phase shift, 
$\phi\rightarrow\phi+2\pi$, the ABS in Eq.(\ref{Eq.12}) clearly has a $4\pi$ period. 
Using Eq.~(\ref{Eq.12}), the corresponding Josephson current becomes,
$I_\pm(\delta\phi) =(e\Delta_S/\hbar)\sum_{n}\partial\epsilon_{\pm,n}(\delta\phi)/\partial\delta\phi=\pm I_c\sin(\delta\phi/2)$,
where $I_c=(e\Delta_S/2\hbar)\sum_{n}\sqrt{\tau_n}$ is the critical value of the Josephson current. In the absence of perturbations 
which violate fermion parity conservation, such a form of fractional Josephson current is a signature of 2DMS on the surface of 3DDS~\cite{Fu2009PRB}. 

\begin{figure}[t]
\includegraphics[scale=0.2]{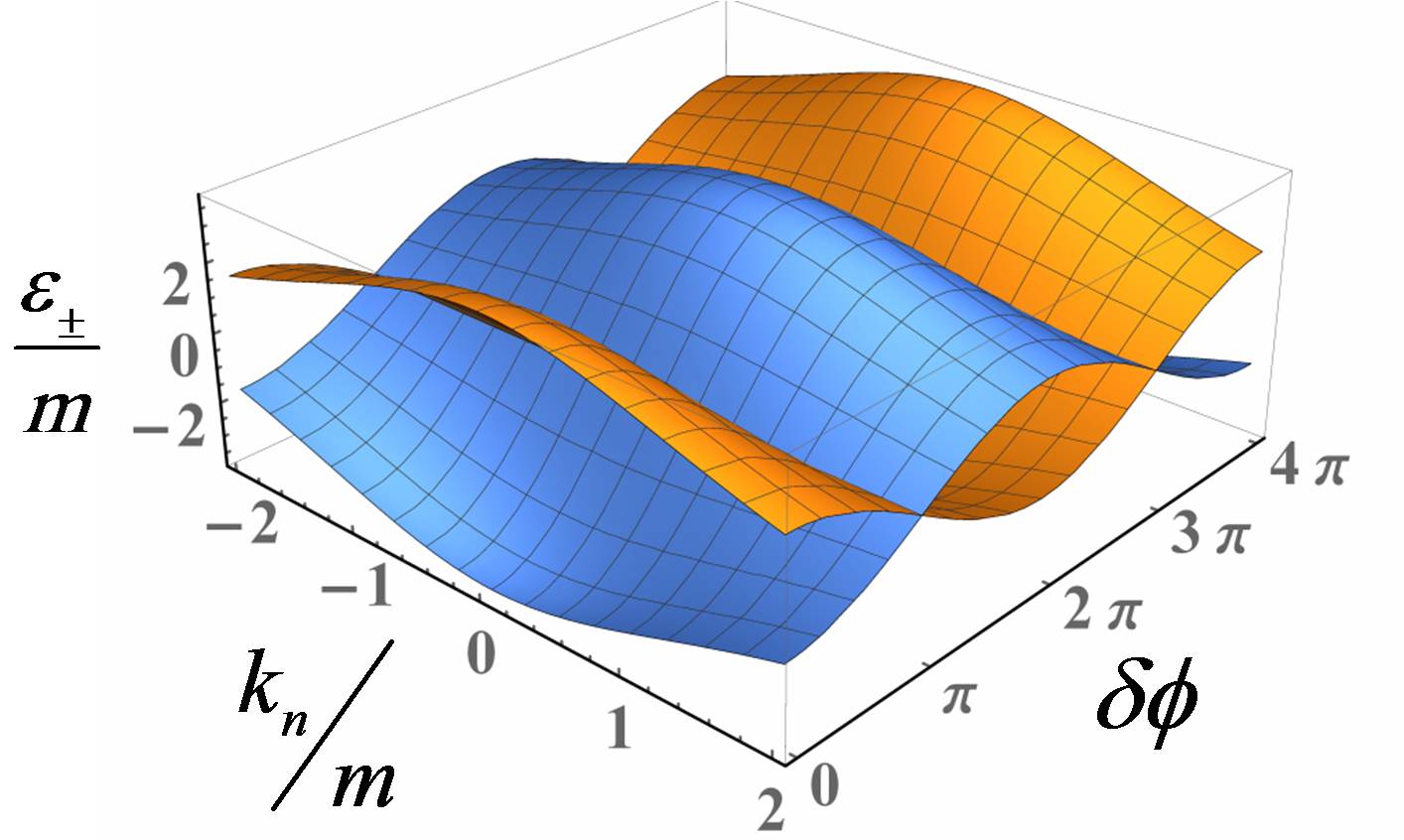}
\caption{(Color online) The Andreev bound states obtained for  Josephson junction of part (b) in Fig.(\ref{Fig.structure}). These states show $4\pi$ periodicity with respect to $\delta\phi$ which is an another direct signature of 2DMS. }
\end{figure}

To summarize we have shown that a 2DMS can be obtained from
singlet pairing as well, provided the superconducting order is pseudo-scalar, i.e.
it changes sign under mirror reflection. In this extent, the present system is a
singlet cousin in the family of odd-parity superconductors~\cite{Fu2010PRL}. 
Our scenario for Majorana surface states does not require proximity to any magnet,
as no triplet pairing is involved here. Perfect Andreev reflection and fractional
Josephson current as two hallmarks of the ensuing Majorana sea leave clear transport
footprints. The junction between such a superconductor and conventional $s$ and $d$ wave
superconductors similar to the odd-parity p-wave superconductors may provide anomalous
flux quantization in units of $h/4e$ ~\cite{Fu2010PRL}. 
Our analysis depends on the existence of the $\gamma^5$ matrix which is a unique
opportunity in even space-time dimensions. Therefore in 1+1 dimension as well, 
Majorana bound states are expected from a competition between the Dirac (Peierls) mass
$m$ and $\Delta_5$ {\em singlet} superconductivity. 
Moreover, gauge transformation can place the odd-parity on the Dirac mass.
Therefore for a peculiar Dirac Hamiltonians where the mass term $m_5$ carries the minus sign needed in
making of the odd-parity, Majorana fermions can arise in $m_5|\Delta_0$ interfaces where $\Delta_0$
is a conventional s-wave superconductors. 

{\em Acknowledgment:} SAJ was supported by Alexander von Humboldt fellowship for experienced researchers.

\newpage
\onecolumngrid

\appendix
\begin{center}
{\bf Supplementary material}
\end{center}
In this supplementary material we expand the details of calculations referenced in the main letter text. 

\section{A. Pairing under $\Delta_5$ order parameter}

The wave functions satisfying the Dirac equation 
\begin{equation}
H_D(\mathbf{k})|\psi\rangle=\epsilon |\psi\rangle
\label{E.01}
\end{equation}
is a four component spinor as,
\begin{equation}
H_{D}(\mathbf{k})=\left(
\begin{array}{cccc}
 m-\mu  & 0 & i k_z & i (k_x-i k_y)\\
 0 & m-\mu  & i (k_x+i k_y) & -ik_z \\
 -i k_z & -i (k_x-ik_y) & -m-\mu & 0\\
 -i (k_x+i k_y) & i k_z & 0 & -m-\mu \\
\end{array}
\right)\left(\begin{array}{c}
\phi_\uparrow \\
\phi_\downarrow \\
\chi_\uparrow\\
\chi_\downarrow\\
\end{array}\right)=\epsilon \left(\begin{array}{c}
\phi_\uparrow \\
\phi_\downarrow \\
\chi_\uparrow\\
\chi_\downarrow\\
\end{array}\right)
\label{E.02}
\end{equation}
where $\phi$ and $\chi$ correspond to the conduction and valence band, respectively. Also, $\uparrow (\downarrow)$ correspons 
to the spin directions of carrier with respect to $z$-axis. The hole part of DBdG equation is obtained by operating with the
charge conjugation on the Hamiltonina and its spinor of Eq.(\ref{E.01}). It consists in the complex conjugation operator $K$ and
multiplication by some $4\times 4$ matrxi the form of which depends on the representation of $\gamma$ matrices and in our
case is~\cite{ZeeBookQFT},
\begin{equation}
\mathcal{C}=\left(\begin{array}{cccc}
0 & 0 & 0& -1 \\
0 & 0& 1 & 0 \\
0 & 1 & 0 & 0 \\
-1 & 0& 0& 0 \\
\end{array}\right)K
\label{E.03}
\end{equation}
such that,
\begin{equation}
\mathcal{C}H_D(\mathbf{k})\mathcal{C}^{-1}{\mathcal C}|\psi\rangle=\epsilon{\mathcal C} |\psi\rangle
\label{E.04}
\end{equation}
which can be recast into the form,
\begin{equation}
H_D(-\mathbf{k})|\psi_c\rangle=-\epsilon |\psi_c\rangle
\label{E.05}
\end{equation}
where the charge conjugated state $|\psi_c\rangle$ is explicitely given by~\cite{ZeeBookQFT},
\begin{equation}
|\psi_c\rangle=\mathcal{C}|\psi\rangle=\left(\begin{array}{c}
-\chi_\downarrow^*\\
\chi_\uparrow^* \\
\phi_\downarrow^*\\
-\phi_\uparrow^*
\end{array}\right)
\label{E.06}
\end{equation}

From $|\psi\rangle$ and $|\psi_c\rangle$ we can construct a Nambu spinor $|\mathbf{\Psi}\rangle$
that satisfies DBdG equation in the direct product space of Eqs.~(\ref{E.04}) and~(\ref{E.06}),
\begin{equation}
H_{DBdG}\left|\mathbf{\Psi}\right\rangle=\epsilon\left|\mathbf{\Psi}\right\rangle
\end{equation}
The explicit representation of the DBdG Hamiltonian matrix in the pseudo-scalar pairing channel becomes,
\begin{equation}
\footnotesize
\left(
\begin{array}{cccccccc}
 m-\mu  & 0 & i k_z & i (k_x-i k_y) & 0 & 0 & i \Delta_5 & 0 \\
 0 & m-\mu  & i (k_x+i k_y) & -ik_z & 0 & 0 & 0 & i \Delta_5 \\
 -i k_z & -i (k_x-ik_y) & -m-\mu  & 0 & i\Delta_5 & 0 & 0 & 0 \\
 -i (k_x+i k_y) & i k_z & 0 & -m-\mu  & 0 & i \Delta_5 & 0 & 0 \\
 0 & 0 & -i \Delta_5 & 0 & \mu -m & 0 & i k_z & i (k_x-ik_y) \\
 0 & 0 & 0 & -i \Delta_5 & 0 & \mu -m & i (k_x+ik_y) & -i k_z \\
 -i \Delta_5 & 0 & 0 & 0 & -i k_z & -i (k_x-ik_y) & m+\mu  & 0 \\
 0 & -i \Delta_5 & 0 & 0 & -i (k_x+ik_y) & i k_z & 0 &m+\mu  \\
\end{array}
\right)\left(\begin{array}{c}
\phi_\uparrow \\
\phi_\downarrow \\
\chi_\uparrow \\
\chi_\downarrow \\
-\chi_\downarrow^* \\
\chi_\uparrow ^* \\
\phi_\downarrow^* \\
-\phi_\uparrow^* \\
\end{array}\right)=\epsilon \left(\begin{array}{c}
\phi_\uparrow \\
\phi_\downarrow \\
\chi_\uparrow \\
\chi_\downarrow \\
-\chi_\downarrow^* \\
\chi_\uparrow ^* \\
\phi_\downarrow^* \\
-\phi_\uparrow^* \\
\end{array}\right)
\label{E.07}
\end{equation}

To clarify the total spin of the pairing symmetry let extract the pairing part of
$\left\langle \mathbf{\Psi}\right| H_{DBdG}\left|\mathbf{\Psi}\right\rangle$
which in terms of $\{\phi_i, \chi_i \}$ becomes,
\begin{equation}
\footnotesize
\left(\begin{array}{c}
\phi_\uparrow^* \\
\phi_\downarrow^* \\
\chi_\uparrow^* \\
\chi_\downarrow^* \\
-\chi_\downarrow \\
\chi_\uparrow \\
\phi_\downarrow \\
-\phi_\uparrow \\
\end{array}\right)^T
\left(
\begin{array}{cccccccc}
 0  & 0 & 0 & 0 & 0 & 0 & i \Delta_5 & 0 \\
 0 & 0  & 0 & 0 & 0 & 0 & 0 & i \Delta_5 \\
 0 & 0 & 0  & 0 & i\Delta_5 & 0 & 0 & 0 \\
 0 & 0 & 0 & 0 & 0 & i \Delta_5 & 0 & 0 \\
 0 & 0 & -i \Delta_5 & 0 & 0 & 0 & 0 & 0 \\
 0 & 0 & 0 & -i \Delta_5 & 0 & 0 & 0 & 0 \\
 -i \Delta_5 & 0 & 0 & 0 & 0 & 0 & 0 & 0 \\
 0 & -i \Delta_5 & 0 & 0 & 0 & 0 & 0 & 0  \\
\end{array}
\right)\left(\begin{array}{c}
\phi_\uparrow \\
\phi_\downarrow \\
\chi_\uparrow \\
\chi_\downarrow \\
-\chi_\downarrow^* \\
\chi_\uparrow ^* \\
\phi_\downarrow^* \\
-\phi_\uparrow^* \\
\end{array}\right)
\label{E.088}
\end{equation}
and simplifies to,
\begin{equation}
i \Delta_5 \left( \phi_\uparrow^* \phi_\downarrow^*-\phi_\downarrow^*\phi_\uparrow^*+\phi_\uparrow\phi_\downarrow-\phi_\downarrow\phi_\uparrow\right)+
i \Delta_5 \left( \chi_\downarrow^*\chi_\uparrow^*-\chi_\uparrow^*\chi_\downarrow^*+\chi_\downarrow\chi_\uparrow-\chi\uparrow\chi_\downarrow\right)
\end{equation}
Which clearly shows the singlet nature of pairing as is expected from singlet character of a pseudo-scalar pairing. 

\section{B. Calculation of the winding number}
To calculate the topological charge of our Hamiltonian we need to construct the particle-hole ($PH$), time-reversal ($TR$), and 
sublattice ($SL$) transformations in the Nambu space.
\subsection{$B_1$. Symmetry operators}
In the main text we introduces the above three operators the explicit form of which is given by,
\begin{equation}
PH=-i \eta_2\otimes\tau_0\otimes\sigma_2 K=\left(
\begin{array}{cccccccc}
 0 & 0 & 0 & 0 & 0 & i & 0 & 0 \\
 0 & 0 & 0 & 0 & -i & 0 & 0 & 0 \\
 0 & 0 & 0 & 0 & 0 & 0 & 0 & i \\
 0 & 0 & 0 & 0 & 0 & 0 & -i & 0 \\
 0 & -i & 0 & 0 & 0 & 0 & 0 & 0 \\
 i & 0 & 0 & 0 & 0 & 0 & 0 & 0 \\
 0 & 0 & 0 & -i & 0 & 0 & 0 & 0 \\
 0 & 0 & i & 0 & 0 & 0 & 0 & 0 \\
\end{array}
\right)K,
\label{Eq.01}
\end{equation}

\begin{equation}
TR=i \eta_0\otimes\tau_3\otimes\sigma_2 K=\left(
\begin{array}{cccccccc}
 0 & 1 & 0 & 0 & 0 & 0 & 0 & 0 \\
 -1 & 0 & 0 & 0 & 0 & 0 & 0 & 0 \\
 0 & 0 & 0 & -1 & 0 & 0 & 0 & 0 \\
 0 & 0 & 1 & 0 & 0 & 0 & 0 & 0 \\
 0 & 0 & 0 & 0 & 0 & 1 & 0 & 0 \\
 0 & 0 & 0 & 0 & -1 & 0 & 0 & 0 \\
 0 & 0 & 0 & 0 & 0 & 0 & 0 & -1 \\
 0 & 0 & 0 & 0 & 0 & 0 & 1 & 0 \\
\end{array}
\right)K,
\label{Eq.02}
\end{equation}
and
\begin{equation}
SL=\eta_2\otimes\tau_3\otimes\sigma_0=\left(
\begin{array}{cccccccc}
 0 & 0 & 0 & 0 & -i & 0 & 0 & 0 \\
 0 & 0 & 0 & 0 & 0 & -i & 0 & 0 \\
 0 & 0 & 0 & 0 & 0 & 0 & i & 0 \\
 0 & 0 & 0 & 0 & 0 & 0 & 0 & i \\
 i & 0 & 0 & 0 & 0 & 0 & 0 & 0 \\
 0 & i & 0 & 0 & 0 & 0 & 0 & 0 \\
 0 & 0 & -i & 0 & 0 & 0 & 0 & 0 \\
 0 & 0 & 0 & -i & 0 & 0 & 0 & 0 \\
\end{array}
\right),
\label{Eq.03}
\end{equation}
where $\eta_\nu$, $\tau_\nu$ and $\sigma_\nu$ with $\nu=\{0,...,3\}$ are Pauli matrices which act on Nambu, band and spin spaces, 
respectively. Also, $\nu=0$ corresponds to 2 by 2 unit matrix. 

\subsection{$B_2$. Chiral DBdG equation}

A Hamiltonian with chiral symmetry can be rotated to block off-diagonal form. The unitary operator constructed from bases 
of chiral matrix can convert the original DBdG to block off-diagonal matrix (Chiral DBdG). The eigenvalues and eigenvectors of 
Eq.~(\ref{Eq.03}) are,
\begin{equation}
\left(
\begin{array}{ll}
\lambda_1= -1 & u_1=\{0,0,0,-i,0,0,0,1\} \\
\lambda_2= -1 & u_2=\{0,0,-i,0,0,0,1,0\} \\
\lambda_3= -1 & u_3=\{0,i,0,0,0,1,0,0\} \\
\lambda_4= -1 & u_4=\{i,0,0,0,1,0,0,0\} \\
\lambda_5=  1 & u_5=\{0,0,0,i,0,0,0,1\} \\
\lambda_6=  1 & u_6=\{0,0,i,0,0,0,1,0\} \\
\lambda_7=  1 & u_7=\{0,-i,0,0,0,1,0,0\} \\
\lambda_8=  1 & u_8=\{-i,0,0,0,1,0,0,0\} \\
\end{array}
\right),
\label{Eq.04}
\end{equation}
which gives the unitary matrix $U_c$ as,
\begin{equation}
U_c=\left(
\begin{array}{cccccccc}
 0 & 0 & 0 & \frac{i}{\sqrt{2}} & 0 & 0 & 0 & -\frac{i}{\sqrt{2}} \\
 0 & 0 & \frac{i}{\sqrt{2}} & 0 & 0 & 0 & -\frac{i}{\sqrt{2}} & 0 \\
 0 & -\frac{i}{\sqrt{2}} & 0 & 0 & 0 & \frac{i}{\sqrt{2}} & 0 & 0 \\
 -\frac{i}{\sqrt{2}} & 0 & 0 & 0 & \frac{i}{\sqrt{2}} & 0 & 0 & 0 \\
 0 & 0 & 0 & \frac{1}{\sqrt{2}} & 0 & 0 & 0 & \frac{1}{\sqrt{2}} \\
 0 & 0 & \frac{1}{\sqrt{2}} & 0 & 0 & 0 & \frac{1}{\sqrt{2}} & 0 \\
 0 & \frac{1}{\sqrt{2}} & 0 & 0 & 0 & \frac{1}{\sqrt{2}} & 0 & 0 \\
 \frac{1}{\sqrt{2}} & 0 & 0 & 0 & \frac{1}{\sqrt{2}} & 0 & 0 & 0 \\
\end{array}
\right).
\label{Eq.05}
\end{equation}
This transformation brings the DBdG Hamiltonian, Eq.~(\ref{E.07}) to the canonical form,
\begin{equation}
\begin{array}{rl}
U_c^{-1} H_{BD}U_c
& =\left(
\begin{array}{cccccccc}
 0 & 0 & 0 & 0 & m+\mu  & 0 & i k_z-\Delta_5 & k_y-i k_x \\
 0 & 0 & 0 & 0 & 0 & m+\mu  & -i k_x-k_y & -i k_z-\Delta_5 \\
 0 & 0 & 0 & 0 & \Delta_5-i k_z & i k_x-k_y & \mu -m & 0 \\
 0 & 0 & 0 & 0 & i k_x+k_y & i k_z+\Delta_5 & 0 & \mu -m \\
 m+\mu  & 0 & i k_z+\Delta_5 & k_y-i k_x & 0 & 0 & 0 & 0 \\
 0 & m+\mu  & -i k_x-k_y & \Delta_5-i k_z & 0 & 0 & 0 & 0 \\
 -i k_z-\Delta_5 & i k_x-k_y & \mu -m & 0 & 0 & 0 & 0 & 0 \\
 i k_x+k_y & i k_z-\Delta_5 & 0 & \mu -m & 0 & 0 & 0 & 0 \\
\end{array}
\right)
\end{array}
\label{Eq.07}
\end{equation}

\subsection{$B_3$. Projector Matrix}
In the presence of translational invariance, the ground states of 3D Dirac superconductor can be constructed as a 
Fermi sea that gets then gapped out by superconducting pairing which separates the filled and empty states
in 3D Brillouin zone (BZ). This enables to define a projector operator as,
\begin{equation}
P(k)=\sum_{a\in{\rm filled}}\left|u_a (k)\right\rangle \left\langle u_a(k)\right|.
\label{Eq.08}
\end{equation}
To construct the projector operator, at first step, we must calculate the wave functions of Eq.~(\ref{Eq.07}) 
for energy eigenvalues of the filled states which are given by,
\begin{equation}
\epsilon_{1,2}=-\sqrt{k_x^2+k_y^2+k_z^2+(m-\Delta_5)^2},~~~~
\epsilon_{3,4}=-\sqrt{k_x^2+k_y^2+k_z^2+(m+\Delta_5)^2}
\label{Eq.09}
\end{equation} 
The wave functions corresponding to the eigenvalues in Eq.~(\ref{Eq.09}) are,
\begin{equation}
\begin{array}{l}
u_1(k)=\left(-\frac{1}{2},0,\frac{1}{2},0,\frac{1}{2} (\delta_m-i k_m \cos (\theta )),\frac{1}{2} i e^{-i \alpha } k_m \sin (\theta ),\frac{1}{2} (\delta_m-i k_m \cos (\theta )),\frac{1}{2} i e^{-i \alpha } k_m \sin (\theta )\right)^T\\
u_2(k)=\left(0,-\frac{1}{2},0,\frac{1}{2},\frac{1}{2} i e^{i \alpha } k_m \sin (\theta ),\frac{1}{2} (\delta_m+i k_m \cos (\theta )),\frac{1}{2} i e^{i \alpha } k_m \sin (\theta ),\frac{1}{2} (\delta_m+i k_m \cos (\theta ))\right)^T\\
u_3(k)=\left(\frac{1}{2},0,\frac{1}{2},0,-\frac{1}{2} \delta_p+i k_p \cos (\theta )),\frac{1}{2} i e^{-i \alpha } k_p \sin (\theta ),\frac{1}{2} (\delta_p+i k_p \cos (\theta )),\frac{1}{2} (-i) e^{-i \alpha } k_p \sin (\theta )\right)^T\\
u_4(k)=\left(0,\frac{1}{2},0,\frac{1}{2},\frac{1}{2} i e^{i \alpha } k_p \sin (\theta ),-\frac{1}{2} (\delta_p-i k_p \cos (\theta )),\frac{1}{2} (-i) e^{i \alpha } k_p \sin (\theta ),\frac{1}{2} (\delta_p-i k_p \cos (\theta ))\right)^T
\end{array}
\label{Eq.10}
\end{equation}
where we have used the notations,
\begin{equation}
\begin{array}{l}
k=\sqrt{k_x^2+k_y^2+k_z^2}\\
\xi_m=\sqrt{k^2+(m-\Delta_5)^2}\\
\xi_p=\sqrt{k^2+(m+\Delta_5)^2}\\
\delta_m=(m-\Delta_5)/\xi_m\\
\delta_p=(m+\Delta_5)/\xi_p\\
k_p=k/\xi_p\\
k_m=k/\xi_m\\
\end{array}
\label{Eq.11}
\end{equation}
The polar and azimuthal angels in the $k$-space are denoted by $(\theta, \alpha)$. 
The projector operator can be calculated by plugging the wave functions of Eq.~(\ref{Eq.10}) into Eq.~(\ref{Eq.08}) 
which gives,
\begin{equation}
P(k)=\frac{1}{2}\left(
\begin{array}{cc}
\mathbf{1} & q\\
q^\dagger & \mathbf{1}\\
\end{array}\right)
\end{equation}
where $\mathbf{1}$ and $q$-matrix are 4 by 4 matrices. Then the $Q$-matrix is obtained as:
 \begin{equation}
 Q(k)=2P(k)-\mathbf{1}=\left(\begin{array}{cc}
 0 & q \\
 q^\dagger & 0
 \end{array}
 \right)
 \end{equation}
Here $\mathbf{1}$ is a unit matrix with 8 by 8 dimension. 
The $q$-matrix which is the off diagonal block of projector is essential for calculation 
of the topologial index, and is given by,
 \begin{equation}
 \footnotesize
 q(k)=
 \left(
 \begin{array}{cccc}
  -\frac{1}{2} (\delta_m+\delta_p+i(k_m-k_p) \cos \theta ) & \frac{1}{2} i e^{i \alpha } (k_m-k_p) \sin \theta  & \frac{1}{2} (-\delta_m+\delta_p-i (k_m+k_p) \cos \theta ) & \frac{1}{2} i e^{i \alpha } (k_m+k_p) \sin \theta  \\
  \frac{1}{2} (k_m-k_p)i e^{-i \alpha} \sin \theta  & \frac{1}{2} (-\delta m-\delta p+i (k_m-k_p) \cos \theta ) & \frac{1}{2} (k_m+k_p) i e^{-i\alpha} \sin \theta  & \frac{1}{2} (-\delta m+\delta p+i (k_m+k_p) \cos \theta ) \\
  \frac{1}{2} (\delta m-\delta p+i (k_m+k_p) \cos \theta ) & -\frac{1}{2} i e^{i \alpha } (k_m+k_p) \sin \theta  & \frac{1}{2} (\delta m+\delta p+i (k_m-k_p) \cos \theta ) & -\frac{1}{2} i e^{i \alpha } (k_m-k_p) \sin \theta  \\
  -\frac{1}{2} i e^{-i \alpha } (k_m+k_p) \sin \theta  & \frac{1}{2} (\delta m-\delta p-i (k_m+k_p) \cos \theta ) & -\frac{1}{2} i e^{-i \alpha } (k_m-k_p) \sin \theta  & \frac{1}{2} (\delta m+\delta p-i (k_m-k_p) \cos \theta ) \\
 \end{array}
 \right)
 \label{Eq.14}
 \end{equation}

\subsection{$B_4$. Topological invariant}
We are now set to calculate the topological invariant from $q(k)$ matrix that is given by the integral
 \begin{equation}
 \mathcal{Q}\left[q\right]=\int \frac{d^3k}{24 \pi^2}\epsilon^{\mu\nu\rho}tr\left[\left(q^{-1}\partial_\mu q\right)\left(q^{-1}\partial_\nu q\right)\left(q^{-1}\partial_\rho q\right)\right],
 \label{Eq.15}
 \end{equation}
Here the   $\epsilon^{\mu\nu\rho}$ is the Levi-Civita tensor and $\mu, \nu, \rho$ run over 3 directions $\{k_x, k_y, k_z\}$. 
By plugging Eq.~(\ref{Eq.14}) into Eq.~(\ref{Eq.15}) we find,
\begin{equation}
\mathcal{Q}=\frac{1}{2}\left(\frac{m+\Delta_5}{\left|m+\Delta_5\right|}-\frac{m-\Delta_5}{\left|m-\Delta_5\right|}\right).
\end{equation} 
For 3D Dirac materials (non superconducting) case where $(\Delta_5=0, m\neq 0$, the winding number will be $\mathcal{Q}=0$ 
while for 3D Dirac superconductor $(\Delta_5\neq 0, m=0)$ the Winding number will be $\mathcal{Q}={\rm sign}(\Delta_5)$.
The above topological index therefore shows a clear change by movign from $\Delta_5$ dominated gapped state to $m$ dominated
gapped state in a $m|\Delta_5$ region. 

\section{C. Transport}
In this section we obtain the transport coefficients in normal and Andreev channels. 
\subsection{$C_1$. Wave functions of 3DDM region}
In 3DDM region, the superconducting gap is zero $\Delta_5=0$, which reduces the Dirac-Bogoliubov-deGennes to,
\begin{equation}
\left( \begin{array}{cc}
\mathcal{H}(\textbf{k})-\mu & 0\\
 0 & \mu+\mathcal{C}\mathcal{H}(\textbf{k})\mathcal{C}^{-1} \\
\end{array}\right)
\left( \begin{array}{c}
u \\
v\\
\end{array}\right)= \epsilon \left(\begin{array}{c}
u \\
v \\
\end{array} \right) 
\label{3DDMBdG}  
\end{equation} 
The electron part of Eq.~(\ref{3DDMBdG}) is completely decoupled from the hole one. Each part has four components 
which can be defined as right or left mover with respect to $z$-axis. Also, each states has two spin orientations, 
$\{\uparrow, \downarrow\}$. For electron part with wave vector $k=|\mathbf{k}|=\sqrt{(\epsilon+\mu)^2-m^2}$, 
we can define the longitudinal and transverse components of wave vector as:
\begin{equation}
\begin{cases}
k_z=|\mathbf{k}|\cos\theta\\
k_{n}=\sqrt{k_x^2+k_y^2}=|\mathbf{k}|\sin\theta\\
\end{cases}
\end{equation}
As mentioned in the main text, the transverse components of wave vector are quantized due to the finite width of junction 
and are labaled by $n=(n_x, n_y)$. So the wave functions can be written as,
\begin{equation}
\begin{array}{l}
\Psi_{e,\uparrow}^{M,\pm}=\left(
\begin{array}{ccccc}
1, & 0,& \mp i \eta_e\cos\theta,& -i \eta_e e^{i \alpha} \sin\theta, & \mathbf{0}^4
\end{array}\right)^Te^{\pm i k_zz+ i \vec{k}_{n}.\vec{r}_{||}},\\
\Psi_{e,\downarrow}^{M,\pm}=\left(
\begin{array}{ccccc}
0, & 1,& -i \eta_e e^{-i \alpha} \sin\theta, &  \pm i \eta_e\cos\theta, & \mathbf{0}^4
\end{array}\right)^Te^{\pm i k_zz+ i \vec{k}_{n}.\vec{r}_{||}},\\
\end{array}
\label{WaveFunctions1}
\end{equation}
where $\mathbf{0}^4$, is the $1\times4$ vector with zero entities, and $\eta_e=\sqrt{(\epsilon+\mu-m)/(\epsilon+\mu+m)}$. 
In order to conserve thei current, the wave functions of Eq.~(\ref{WaveFunctions1}) need the normalization factor, 
$1/\sqrt{2\eta_e \cos\theta}$. For the wave functions of hole part we obtain,
\begin{equation}
\begin{array}{l}
\Psi_{h,\uparrow}^{M,\pm}=\left(
\begin{array}{ccccc}
\mathbf{0}^4,& i \eta_h e^{-i \alpha}\sin\theta', &  \mp i \eta_h \cos\theta', & 0, & 1\\
\end{array}\right)^Te^{\pm i k_z'z+i \vec{k}_{n}.\vec{r}_{||}},\\
\Psi_{h,\downarrow}^{M,\pm}=\left(
\begin{array}{ccccc}
\mathbf{0}^4,&  \pm i \eta_h \cos\theta', &  i \eta_h e^{i \alpha}\sin\theta', & 1, & 0
\end{array}\right)^Te^{\pm i k_z'z+ i \vec{k}_{n}.\vec{r}_{||}},
\end{array}
\label{WaveFunction2}
\end{equation}
where $\eta_h=\sqrt{(\epsilon-\mu-m)/(\epsilon-\mu+m)}$ and $k'=|\mathbf{k'}|=\sqrt{(\epsilon-\mu)^2-m^2}$. The polar angle of propagation for hole-like quasi-particle is $\cos\theta'=k_{n}/k'$. Owing to translation invariance in the transverse directions, $k_{||}$ is conserved at the scattering processes. The $k_z'$ can be written as,
\begin{equation}
k_z'=\sqrt{(\epsilon-\mu)^2-m^2)-k_{n}^2}.
\end{equation}
This equation implies that for incident angle ($\theta$) beyond critical value of,
\begin{equation}
\theta_c=\arcsin\left((\sqrt{\frac{(\epsilon-\mu)^2-m^2}{(\epsilon+\mu)^2-m^2}}\right),
\end{equation}
the perpendicular component of the wave vector for hole-like wave functions will be imaginary. 
This means hole-like states beyond these propagation angles can not contribute to transport. 
The normalization factor for wave functions of Eq.~(\ref{WaveFunction2}) is $1/\sqrt{2\eta_h\cos\theta'}$.

\subsection{$C_2$. Wave functions of 3DDS region}
In superconducting region the pairing matrix $\hat{\Delta}_S=\Delta_5\gamma^5$ 
and we have $\mu_S\gg m$. In this regime the DBdG Hamiltonian will be:
\begin{equation}
\left( \begin{array}{cc}
\mathcal{H}(\textbf{k})-\mu_S & \hat{\Delta}_S \gamma_0 e^{i \phi}\\
\gamma_0\hat{\Delta}^\dagger_S e^{-i \phi} & \mu_S+\mathcal{C}\mathcal{H}(\textbf{k})\mathcal{C}^{-1} \\
\end{array}\right)
\left( \begin{array}{c}
u \\
v\\
\end{array}\right)= \epsilon \left(\begin{array}{c}
u \\
v \\
\end{array} \right).
\label{Eq.DBdG}  
\end{equation}
The eigenvalues of Eq.(\ref{Eq.DBdG}) are,
\begin{equation}
\epsilon= \pm \sqrt{(|\mathbf{k}| \pm \mu_S)^2+\Delta_5^2}.
\label{Eq.Seigenvalues}
\end{equation}
Each eigenvalues are doubly degenerate because of spin. The wave vectors of electron-like and hole-like quasi-particles 
corresponding to the eigenvalues of Eq.(\ref{Eq.Seigenvalues}) are,
\begin{equation}
\begin{cases}
|\mathbf{k}_e^S|=\left(\mu_S+\sqrt{\epsilon^2-\Delta_S^2}\right)/\hbar v_F\\
|\mathbf{k}_h^S|=\left(\mu_S-\sqrt{\epsilon^2-\Delta_S^2}\right)/\hbar v_F\\
\end{cases}
\end{equation}
with corresponding wave functions:
\begin{equation}
\begin{array}{l}
\Psi_{e, \kappa=1}^{S, \pm}=
\left( \begin{array}{cccccccc}
1, & 0, & \mp i\cos\theta_S, & -ie^{i\alpha}\sin\theta_S, & \mp e^{-i\phi-i\beta}\cos\theta_S, & -e^{-i\phi +i\alpha-i\beta}\sin\theta_S,& -i e^{-i \phi-i \beta}, & 0
\end{array}\right)^T e^{\pm i k_z^Sz}\\
\Psi_{e, \kappa=2}^{S, \pm}=
\left( \begin{array}{cccccccc}
0, & 1, & -ie^{-i\alpha}\sin\theta_S, & \pm i \cos\theta_S, & - e^{-i\phi-i\alpha-i\beta}\sin\theta_S, &  \pm e^{-i\phi-i\beta}\cos\theta_S,& 0, & -i e^{-i\phi-i\beta} 
\end{array}\right)^T e^{\pm i k_z^Sz}\\
\Psi_{h, \kappa=1}^{S, \pm}=
\left( \begin{array}{cccccccc}
\mp i \cos\theta_S, & i e^{i\alpha}\sin\theta_S, & 1, & 0, & -i e^{i \beta-i \phi}, &   0,& \mp  e^{-i\phi+i\beta}\cos\theta_S, &  e^{i\alpha+i\beta-i\phi}\sin\theta_S 
\end{array}\right)^T e^{\pm i k_z'^Sz}\\
\Psi_{h, \kappa=2}^{S, \pm}=
\left( \begin{array}{cccccccc}
i e^{-i\alpha}\sin\theta_S, & \pm i \cos\theta_S, & 0, & 1, & 0, &   -i e^{-i\phi+i\beta}, &  e^{-i\alpha-i\phi+i\beta}\sin\theta_S, & \pm e^{-i\alpha+i\beta-i\phi}\sin\theta_S 
\end{array}\right)^T e^{\pm i k_z'^Sz}
\end{array}
\label{Wavefunctions3}
\end{equation}
Note that all wave functions in Eq.~(\ref{Wavefunctions3}) must be multiplied by a the Bloch phase of parallel component 
of wave vector,  $\exp(i \mathbf{k}_{n}.\mathbf{r}_{||})$ and the normalization factor $1/2\sqrt{\cos\theta_S}$.

\subsection{$C_3$. Boundary condition}
Using wave functions of Eqs.~(\ref{WaveFunctions1}),~(\ref{WaveFunction2} and~(\ref{Wavefunctions3}), 
one can solve the boundary conditions of Eq.~(4) in the main text to complete the construction of scattering states. 
The reflection amplitudes for an $\uparrow$-spin incident electron are derived as,
\begin{equation}
\begin{array}{l}
r_{N,\uparrow}=\frac{\left(\omega_e^{c^2}-\omega_e^{s^2}-1 \right)\left(\chi_h^{+}\cos 2\beta+2i \omega_h^{c}\sin 2\beta\right)-\left(\chi_h^{-}\left(\omega_e^{c^2}-\omega_e^{s^2}+1 \right)+4 \omega_e^s\omega_h^s\right)}{\left(\chi_e^{+}\chi_h^{+}+4\omega_e^c\omega_c^c\right)\cos 2\beta+2i\left(\chi_h^{+}\omega_e^c+\chi_e^{+}\omega_h^c\right)\sin 2\beta-\left(\chi_e^{-}\chi_h^{-}-4\omega_e^s\omega_h^s\right)},\\
\\
r_{N,\downarrow}=\frac{2e^{i\alpha}\omega_e^c\{\left(\omega_e^s(\chi_h^{-}-\chi_h^{+}\cos 2\beta)-2i\omega_e^s\omega_h^s\sin 2\beta-2 \omega_h^s\right)\}}{\left(\chi_e^{+}\chi_h^{+}+4\omega_e^c\omega_c^c\right)\cos 2\beta+2i\left(\chi_h^{+}\omega_e^c+\chi_e^{+}\omega_h^c\right)\sin 2\beta-\left(\chi_e^{-}\chi_h^{-}-4\omega_e^s\omega_h^s\right)},\\
\\
r_{A,\uparrow}=\frac{2i e^{i\alpha-i\phi}\sqrt{\omega_e^c\omega_h^c}\left(2(\omega_e^s+\omega_h^s)\cos \beta+2i(\omega_e^s\omega_h^c+\omega_e^c\omega_h^s)\sin\beta\right)}{\left(\chi_e^{+}\chi_h^{+}+4\omega_e^c\omega_c^c\right)\cos 2\beta+2i\left(\chi_h^{+}\omega_e^c+\chi_e^{+}\omega_h^c\right)\sin 2\beta-\left(\chi_e^{-}\chi_h^{-}-4\omega_e^s\omega_h^s\right)},\\
\\
r_{A,\downarrow}=\frac{-4i e^{-i\phi}\sqrt{\omega_e^c\omega_h^c}\left((\omega_e^c+\omega_h^c)\cos\beta-i(1+\omega_e^c\omega_h^c-\omega_e^s\omega_h^s)\sin\beta\right)}{\left(\chi_e^{+}\chi_h^{+}+4\omega_e^c\omega_c^c\right)\cos 2\beta+2i\left(\chi_h^{+}\omega_e^c+\chi_e^{+}\omega_h^c\right)\sin 2\beta-\left(\chi_e^{-}\chi_h^{-}-4\omega_e^s\omega_h^s\right)},
\end{array}
\label{Eq.ReflectionAmplitudes}
\end{equation}
where the notation is:
\begin{equation}
\begin{cases}
\omega_{e(h)}^s=\eta_{e(h)}\sin\theta (\theta')\\
\omega_{e(h)}^c=\eta_{e(h)}\cos\theta (\theta')\\
\chi_{e(h)}^{\pm}=\eta_{e(h)}\pm 1
\end{cases}.
\label{Eq.ProbabilitiesVairables}
\end{equation}
In zero mode regime, $\epsilon \rightarrow 0$, these reflection amplitudes reduces to Eq.~(6) of the main text.

\section{D. Josephson Current}
In order to obtain the supercurrent for $\Delta_5 | m | \Delta_5$ Josephson junction we use transfer matrix method to 
calculate the energy dependence of the Andreev bound states (ABS)\cite{Titov2006PRB}. In the short junction limit, 
$L \ll \xi_S$, when an $\uparrow$-spin electron hits the right interface of $\Delta_5|m|\Delta_ 5$ junction at $z=L$,
the reflection amplitudes of Eq.~(\ref{Eq.ReflectionAmplitudes}) reduce to:
\begin{equation}
\begin{cases}
r_{N,\uparrow\uparrow}^{(1)}=\frac{\cos (\beta ) \left(\eta ^2 \cos (2 \theta )-1\right)}{\left(\eta ^2+1\right) \cos (\beta )+2 i \eta  \sin (\beta ) \cos (\theta )}\\
\\

r_{N,\uparrow\downarrow}^{(1)}=-\frac{2 e^{i \alpha } \eta ^2 \cos (\beta ) \sin (\theta ) \cos (\theta )}{\left(\eta ^2+1\right) \cos (\beta )+2 i \eta  \sin (\beta ) \cos (\theta )}\\
\\
r_{A,\uparrow\uparrow}^{(1)}=\frac{2 i \eta  e^{i (\alpha -\phi )} \sin (\theta ) \cos (\theta )}{\left(\eta ^2+1\right) \cos (\beta )+2 i \eta  \sin (\beta ) \cos (\theta )}\\
\\
r_{A,\uparrow\downarrow}^{(1)}=-\frac{2 i \eta  e^{-i \phi } \cos ^2(\theta )}{\left(\eta ^2+1\right) \cos (\beta )+2 i \eta  \sin (\beta ) \cos (\theta )}\\
\\

\end{cases}.
\label{Eq.ShortJunctionFor1}
\end{equation}
Here, $\eta=\sqrt{(\mu-m)/(\mu+m)}$.  We use superscript $(2)$ when an $\downarrow$-spin electron hits the interface. 
Similarly, $(3)$ and $(4)$ are using for $\uparrow$-spin and $\downarrow$-spin hole, respectively. The first index in 
the doulbe spin indices indicates the spin configuration of incoming particle and the second one denotes the outgoing one.
The primarily matrices $M_1$ and $M_2$ can be constructed by these 16 amplitudes as~\cite{Titov2006PRB}:
\begin{eqnarray}
M_1=
\left(
\begin{array}{cccc}
 0 & 0 & -r_{A,\uparrow\uparrow}^{(3)} & -r_{A,\downarrow\uparrow}^{(4)} \\
 0 & 0 & -r_{A,\uparrow\downarrow}^{(3)} & -r_{N,\downarrow\downarrow}^{(4)} \\
 1 & 0 & -r_{N,\uparrow\uparrow}^{(3)} & -r_{N,\downarrow\uparrow}^{(4)} \\
 0 & 1 & -r_{N,\uparrow\downarrow}^{(3)} & -r_{N,\downarrow\downarrow}^{(4)} \\
\end{array}
\right),\\
M_2=\left(
\begin{array}{cccc}
 r_{N,\uparrow\uparrow}^{(1)} & r_{N,\downarrow\uparrow}^{(2)} & -1 & 0 \\
 r_{N,\uparrow\downarrow}^{(1)} & r_{N,\downarrow\downarrow}^{(2)} & 0 & -1 \\
 r_{A,\uparrow\uparrow}^{(1)} & r_{A,\downarrow\uparrow}^{(2)} & 0 & 0 \\
 r_{A,\uparrow\downarrow}^{(1)} & r_{A,\downarrow\downarrow}^{(2)} & 0 & 0 \\
\end{array}
\right).
\label{Eq.PrimarilyTransferMatrix}
\end{eqnarray}
\begin{figure}
\includegraphics[scale=0.2]{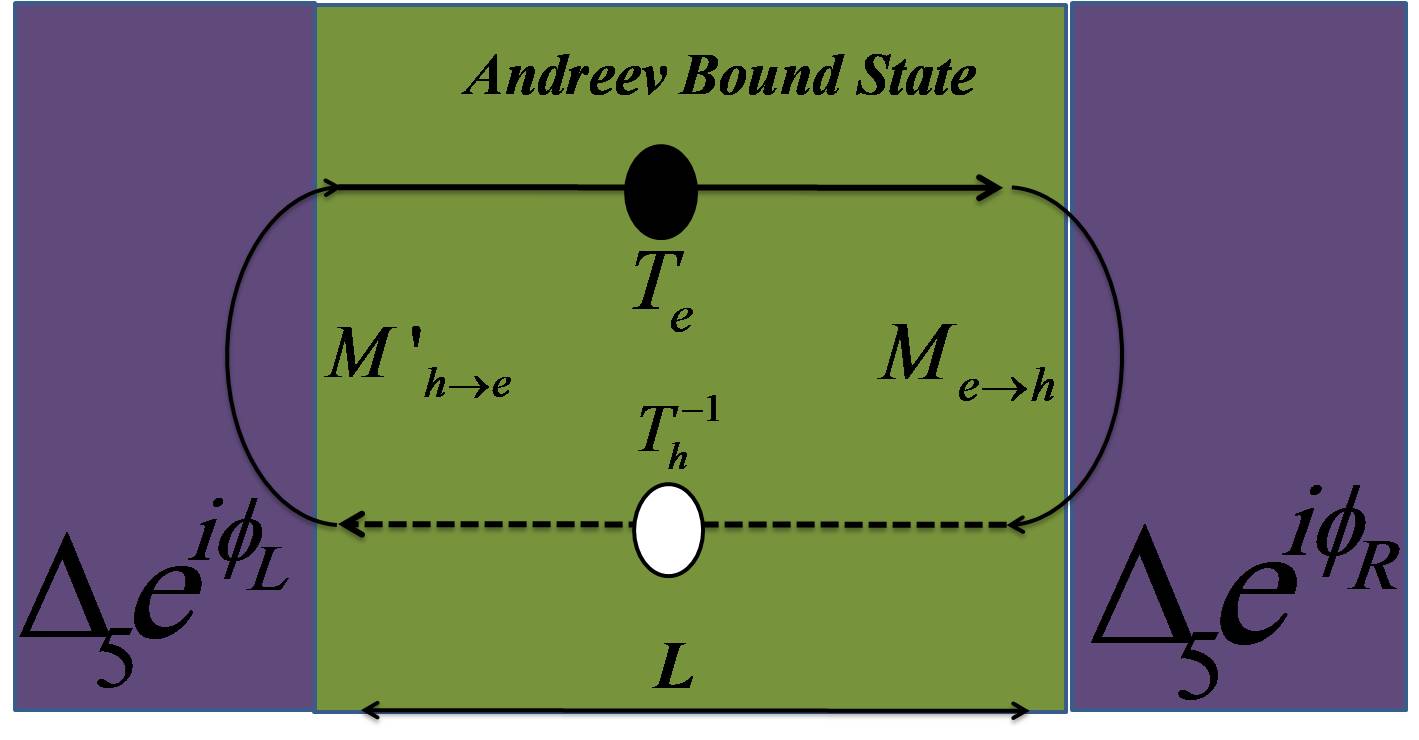}
\caption{The $\Delta_5 | m | \Delta_5$ Josephson junction. The Andreev bound states can be constructed using the combination of transfer matrices $\{T_e, M_{e\rightarrow h}, T_h^{-1}, M_{h\rightarrow e}\}$, }
\end{figure}
Using these two matrices, the transfer matrix at the right interface which convert electron to hole wave functions can be calculated as:
\begin{equation}
M_{e\rightarrow h}=M_1^{-1}M_2=\left(\begin{array}{cccc}
M_{11} & M_{12} & M_{13} & M_{14}\\
M_{21} & M_{22} & M_{23} & M_{24}\\
M_{31} & M_{32} & M_{33} & M_{34}\\
M_{41} & M_{42} & M_{43} & M_{44}\\
\end{array} \right)
\label{Eq.MeToh}
\end{equation}
With the similar procedure, one can construct the transfer matrix in the left interface which can convect hole to 
electron wave function, $M'_{h\rightarrow e}$. Also, there are two another transfer matrices, $\{T_e, T_h\}$, which 
can be used for transferring electron or hole wave function from left interface to the right one. The energy of 
ABSs is then obtained from the condition,
\begin{equation}
\det[\mathbf{1}-M'_{h\rightarrow e}T_h^{-1}M_{e\rightarrow h}T_e]=0.
\label{Eq.ABS01}
\end{equation}  
After some straightforward algebra we obtain the energy of ABSs as given in Eq.~(8) of the main text. 
The derivation of Josephson current from ABSs is explained in the main text as well.


\begin{thebibliography}{99}
\bibitem{BernevigBook}
B. Bernevig and T. Hughes, \textit{Topological Insulators and Topological Superconductors} (Princeton University Press, 2013).

\bibitem{Zhang2011RMP}
X.-L. Qi and S.-C. Zhang, Rev. Mod. Phys. {\bf 83}, 1057
(2011).

\bibitem{ShenBook2013}
S. Shen, {\it Topological Insulators: Dirac Equation in Condensed Matters} (Springer London, Limited, 2013).

\bibitem{Alicea}
J. Alicea, Rep. Prog. Phys. {\bf 75}, 076501 (2012).

\bibitem{Beenakker2013rev}
C. W. J. Beenakker, Ann. Rev. Cond. Matt. \textbf{4}, 113
(2013).

\bibitem{Fu2008PRL}
L. Fu and C. L. Kane, Phys. Rev. Lett. \textbf{100}, 096407
(2008).

\bibitem{Fu2009PRB}
L. Fu and C. L. Kane, Phys. Rev. B. \textbf{79}, 161408 (2009).

\bibitem{Sau}
J. D. Sau, R. M. Lutchyn, S. Tewari, and S. Das Sarma,
Phys. Rev. Lett. \textbf{104}, 040502 (2010).

\bibitem{vonOppen}
Y. Oreg, G. Refael, and F. von Oppen, Phys. Rev. Lett.
\textbf{105}, 177002 (2010).

\bibitem{Romito}
A. Romito, J. Alicea, G. Refael, and F. von Oppen,
Phys. Rev. B \textbf{85}, 020502 (2012).

\bibitem{Stone2004}
M. Stone and R. Roy, Phys. Rev. B \textbf{69}, 184511 (2004).

\bibitem{Fuseya2011PRL}
Y. Fuseya, M. Ogata, and H. Fukuyama, Phys. Rev.
Lett. \textbf{102}, 066601 (2009).

\bibitem{Zhang2009NP}
H. Zhang, C.-X. Liu, X.-L. Qi, X. Dai, Z. Fang, and
S.-C. Zhang, Nat. Phys. \textbf{5}, 438 (2009).

\bibitem{Narayan2014PRL}
A. Narayan, D. Di Sante, S. Picozzi, and S. Sanvito,
Phys. Rev. Lett. \textbf{113}, 256403 (2014).

\bibitem{Liu2014Science}
Z. K. Liu, B. Zhou, Y. Zhang, Z. J. Wang, H. M. Weng,
D. Prabhakaran, S.-K. Mo, Z. X. Shen, Z. Fang, X. Dai,
Z. Hussain, and Y. L. Chen, Science \textbf{343}, 864 (2014).

\bibitem{ZeeBookQFT}
A. Zee, \textit{Quantum Field Theory in a Nutshell:} (Second Edition) (Princeton University Press, 2010).

\bibitem{Capelle1999PRB2}
K. Capelle and E. K. U. Gross, Phys. Rev. B. \textbf{59}, 7140 (1999).

\bibitem{Supplement}
\textit{Please see supplementary article.}

\bibitem{Schnyder2008PRB}
A. P. Schnyder, S. Ryu, A. Furusaki, and A. W. W.
Ludwig, Phys. Rev. B. \textbf{78}, 195125 (2008).

\bibitem{Chiu2016RMP}
C.-K. Chiu, J. C. Y. Teo, A. P. Schnyder, and S. Ryu,
Rev. Mod. Phys. \textbf{88}, 035005 (2016).

\bibitem{Andreev1964JETP}
A. Andreev, Sov. Phys. JETP \textbf{19}, 1228 (1964).

\bibitem{Salehi2015AOP}
M. Salehi and S. A. Jafari, Ann. Phys. \textbf{359}, 64 (2015).

\bibitem{Knez2012PRL}
I. Knez, R.-R. Du, and G. Sullivan, Phys. Rev. Lett.
\textbf{109}, 186603 (2012).

\bibitem{Blonder1982PRB}
G. E. Blonder, M. Tinkham, and T. M. Klapwijk, Phys.
Rev. B. \textbf{25}, 4515 (1982).

\bibitem{Fu2010PRL}
L. Fu and E. Berg, Phys. Rev. Lett. \textbf{105}, 097001 (2010).


\end{thebibliography}

\begin{thebibliography}{99}

\bibitem{Titov2006PRB}
M. Titov and C. W. J. Beenakker, Phys. Rev. B. \textbf{74}, 041401 (2006).

\end{thebibliography}
\end{document}